\documentclass[10pt,preprint]{aastex}
\usepackage{natbib}
\usepackage{amsmath}
\begin{document}
\newcommand{\crc}{cells$/r_c$}
\newcommand{\rr}[1]{$R_{#1}$}

\shorttitle{Self-Convergence of Radiatively Cooling Clumps}
\shortauthors{Yirak et al.}

\title{Self-Convergence of Radiatively Cooling Clumps}

\author{Kristopher Yirak\altaffilmark{1}, Adam Frank\altaffilmark{1}, Andrew J.\
 Cunningham\altaffilmark{1,2}}
\altaffiltext{1}{Department of Physics and Astronomy, University of Rochester, \
Rochester, NY 14620 \\Email contact: yirak@pas.rochester.edu}
\altaffiltext{2}{Lawrence Livermore National Laboratory, Livermore, CA 94551}

\begin{abstract}
Isolated regions of higher density populate the ISM on all scales -- from molecular clouds, to the star-forming regions known as cores, to  heterogeneous ejecta found near planetary nebulae and supernova remnants. These {\it clumps} interact with winds and shocks from nearby energetic sources. Understanding the interactions of shocked clumps is vital to our understanding of the composition, morphology, and evolution of the ISM. The evolution of shocked clumps are well-understood in the limiting ``adiabatic'' case where physical processes such as self-gravity, heat conduction, radiative cooling, and magnetic fields are ignored.  In this paper we address the issue of evolution and convergence when one of these processes - radiative cooling - is included.  

Numeric convergence studies demonstrate that the evolution of an adiabatic clump is well-captured by roughly 100 cells per clump radius. The presence of radiative cooling, however, imposes limits on the problem due to the removal of thermal energy. Numerical studies which include radiative cooling typically adopt the 100--200 cells per clump radius resolution. In this paper we present the results of a convergence study for radiatively cooling clumps undertaken over a broad range of resolutions, from 12 to 1,536 cells per clump radius, employing adaptive mesh refinement (AMR) in a 2D axisymmetric geometry ({\it 2.5D}). We also provide a fully 3D simulation, at 192 cells per clump radius, which supports our 2.5D results. We find no appreciable self-convergence at $\sim$100 cells per clump radius as small-scale differences owing to increasingly resolving the {\it cooling length} have global effects. We therefore conclude that self-convergence is an insufficient criterion to apply on its own when addressing the question of sufficient resolution for radiatively cooled shocked clump simulations. We suggest the adoption of alternate criteria to support a statement of sufficient resolution, such as the demonstration of adequate resolution of the cooling layers behind shocks. We discuss an associated refinement criteria for AMR codes.
\end{abstract}

\section{Introduction}
The distribution of matter is not uniform across a wide range of scales in the universe. Within our own galaxy the ISM is not smooth and, in particular, density inhomogenities are found in molecular clouds, and within these clouds matter is distributed unevenly all the way down to the star-forming regions known as molecular cloud cores. Clumps of material exist on smaller scales as well. This heterogeneous distribution of matter is required, of course, for star and planet formation. On the other hand, energetic sources such as young stellar objects (YSOs), planetary nebulae (PNe), and supernovae inject kinetic energy back into their environments in the form of winds, jets, and shocks. It becomes a matter of central interest, therefore, to understand how clumps and winds, jets, and shocks interact.

One would expect a wind or shock coming from an astrophysical source to have some curvature. It follows that a shock sweeping over a spherical clump will have a radius of curvature which may or may not be comparable to that of the clump. For analytic simplicity many studies consider the ``small clump'' limit, in which the curvature of the shock is so large compared to the clump that it may be considered planar \citep[see e.g.][hereafter KMC94]{klein1994}. Studies where this limit is not applied include those addressing questions such as the evolution of a heterogeneous envelope surrounding PNe \citep{steffen2004}, or energetic jets impinging on a single clump \citep[e.g.][]{raga2002hh110, fragile2004} or a host of clumps \citep{yirak2008}.

Initial studies of single clump/shock interaction focused on the early stages of the interaction, where the solution remained amendable to linear approximations. The evolution late in time, when the behavior becomes highly nonlinear, remains intractable from a purely analytic standpoint and therefore has benefitted greatly from numerical investigation --- a review of the pioneering literature may be found in KMC94 or \cite{poludnenko2002}. Illustrating the maturity of the field, a variety of physical processes have been included in the studies. KMC94 discussed systematically the evolution of a single, adiabatic, nonmagnetized, non-thermally conducting shocked clump overrun by a planar shock in axisymmetry (``2.5D''). Similar simulations were carried out in three dimensions (3D) by \cite{stone1992}. Magnetic fields can play an important part in the clump evolution --- the role of field orientation was first investigated by \cite{jones1996} and more recently by \cite{vanloo2007,shin2008}. Smooth cloud boundaries \citep[e.g.][]{nakamura2006}, and systems of clumps \citep[e.g.][]{poludnenko2002} also have been studied. The similar problem involving clump-clump collisions has also received attention \citep{miniati1997, kleinwoods1998}. Studies predominantly use an Eulerian mesh with a single- or two-fluid method to solve the inviscid Euler equations. One notable exception is \cite{pittard2009}, who use a ``$\kappa-\epsilon$'' model to explicitly model the effects of sub-grid-scale turbulent viscosity by adding turbulence-specific viscosity and diffusion terms to the fluid equations.

The role of radiative cooling has also been studied \citep[e.g.][]{mellema2002, fragile2004, orlando2005}. The energy loss associated with optically thin radiative cooling significantly modifies the evolution of the shocked clump. In the adiabatic regime, the shocked clump collapses initially before expanding in the direction perpendicular to the shock due to the increased thermal energy. For strong cooling, \cite{mellema2002,fragile2004} find that the clump not only does not reexpand but in fact fragments and collapses down to grid scales. \cite{orlando2005}, who consider the role of thermal conduction as well as magnetic fields and radiative losses, find that even for weaker cooling the evolution differs from the adiabatic case. There exists a broad range of velocities, temperatures, and densities that admit radiative cooling. Radiative cooling, therefore, is a significant effect which requires careful investigation.

As with any numerical investigation the appropriate choice of spatial resolution is of critical importance. KMC94 suggested that $\sim$100 cells per clump radius was required to adequately model the interaction. Authors subsequent to KMC94 typically have adopted resolutions in the range of 100-200 cells per clump radius when performing simulations in 2- or 2.5D. Three-dimensional simulations being more computationally expensive, authors in some cases choose lower resolution. A small 3D resolution study in \cite{agertz2007}, for example, investigated the behavior of a 3D adiabatic shocked clump with $\sim$6, $\sim$12, and $\sim$25 cells per clump radius. In contrast, \cite{shin2008} employed 120 cells per clump radius to model the effects of imposed magnetic fields on a single adiabatic shocked clump in 3D. \cite{orlando2005} also addressed resolution by considering two 3D simulations, at 105 and 132 cells per clump radius, respectively, and found good agreement between the two.

Is the choice of 100-200 cells per clump radius always appropriate in clump simulations when different physical processes are included? \cite{kleinwoods1998} investigated clump-clump collisions in order to better understand the nonlinear thin shell instability (NTSI). The NTSI can result at the interface between two opposing flows, causing any perturbation at the interface to grow. When modelling adiabatic or radiatively cooling clumps, they found 128 cells per clump radius to be sufficient. When modelling isothermal clumps, however, they found that at this resolution the NTSI was damped, and resolutions 4-5 times higher were required to model the instability. The cause was the vanishing of the thin interaction layer behind isothermal shocks, which was reduced down to at most a few cells. This, combined with the specifics of their numerical scheme, resulted in an artificial numerical damping of the instability.  Thus \cite{kleinwoods1998}  were required to increase the resolution to counteract this effect. Such a requirement of adequate resolution applies not only to clump-clump collisions, of course. In simulations of environments where self-gravity plays a role, for example, typically the resolution requirements are described in terms of adequate resolution of the local Jeans length \citep[see e.g.][]{truelove1998}. It is always important that the relevant physics on the scales of interest be adequately resolved. 

This resolution/scale rule holds true for radiative cooling. The length scale of interest for radiative cooling behind shocks is the {\it cooling length}: the distance behind the radiative shock where the shocked material cools according to some cooling function $\propto n^2\Lambda(T)$. As will be demonstrated, such a scale may not be adequately resolved in all cases previously considered. On the one hand, when not fully resolved, the cooling length will be artifically increased, implying that material a given distance behind a shock is at a higher temperature than it should be. On the other hand, if the cooling length is highly underresolved, the cooling will take place on of order the cell size, and the simulation will be essentially isothermal \citep{dysonwilliams1997}. If, as is the case in shocked clump interactions, multiple shocks interact with each other, this underresolution may lead to global incorrect results.

In contrast to studies of adiabatic clumps, there has not been a convergence study undertaken for cooling clumps. A possible reason for this is that different cooling tables or multiphysics are employed in different studies, so the fine details may not be generalizeable. However, it remains worthwhile to study questions pertaining to cooling length as a necessary criterion for the resolution of radiatively cooling shocked clump simulations.

We therefore seek to understand the effect of varying resolution in radiatively cooling shocked clump simulations. In particular, we address a sequence of linked questions: 1) how do the simulations vary, qualitatively and quantitatively, with resolution; 2) do the simulations converge in the same manner as adiabatic clump models; 3) since these simulations are performed in 2.5D, how might the conclusions extend to fully 3D simulations; and 4) what conclusions may be drawn in general about simulations of radiatively cooled clumps?

In \S~\ref{problem} we describe the problem and physical scales of interest, and \S~\ref{numerics} discusses our numerical methods. \S~\ref{results} highlights the results from the investigation, which are discussed in more quantitative detail in \S~\ref{analysis}. We conclude in \S~\ref{discussion}.

\section{Description of the Problem}\label{problem}
\subsection{Evolution of a shocked clump}\label{prob_adiabatic}
The evolution of an adiabatic shocked clump progresses through several stages. The external shock will sweep across or ``pass over'' the clump radius in a time $t_{sp}=r_c/v_s$, where $r_c$ is the radius of the clump and $v_s$ the shock velocity. When the shock hits the clump, a new shock wave is transmitted into the clump. This transmitted shock crosses the clump in a time known as the {\it cloud-crushing time} $t_{cc}$ (KMC94),

\begin{equation}\label{eq_tcc}
  t_{cc} = \frac{r_c}{v_{s,c}}\simeq\frac{\chi^{1/2}r_c}{v_s}
\end{equation}
where $v_{s,c}$ is the transmitted (internal) shock velocity and $\chi\equiv\rho_c/\rho_a$, the clump-to-ambient density ratio. The approximation $v_{s,c}\sim v_s/\chi^{1/2}$ comes from considering approximate ram pressure balance for high-Mach number shocks, where $\rho_a v_s^2\simeq \rho_c v_{s,c}^2$. Since $\chi > 1$, then $t_{sp} < t_{cc}$, and the cloud will be enclosed in the hot postshock flow before the transmitted shock crosses the clump. No longer in pressure balance, the clump will begin to be compressed on all sides via new ``transmitted shocks.'' When the transmitted shocks meet and interact, a rarefaction is produced which causes the clump to reexpand. Thereafter, the clump is destroyed and mixes with the postshock flow in a few cloud-crushing times. See KMC94 for a more detailed description on the evolution of an adiabatic shocked clump. How radiative cooling modifies this evolution is discussed in \S~\ref{prob_cooling}.

The two main sources of disruption of the clump at the clump/postshock interface are the Kelvin-Helmholtz (KH) and Rayleigh-Taylor (RT) instabilities. The KH instability will occur due to the relative velocity between the postshock flow and the clump material. The growth rate of the KH instability in the case of $\chi\gg 1$ is $t_\textnormal{KH}=\chi^{1/2}/(k v_{rel})$ \citep{chandrasekhar1961}, where $v_{rel}$ is the relative velocity between the clump and postshock material. If we assume the relative velocity is nearly equal to the external postshock velocity ($v_{ps}=3 v_s/4$ for $M\gg1$), then the KH growth time is

\begin{equation}\label{eq_tkh}
  t_\textnormal{KH} = \frac{4 \chi^{1/2}}{3 k v_s} \sim \frac{t_{cc}}{k r_c}\quad .
\end{equation}
While the smallest wavelength instabilities will grow the fastest, it has been seen that the most disruptive modes are those of order the clump radius \citep[KMC94,][]{poludnenko2002}. Therefore, for the dominant modes for an adiabatic shock clump interaction will occur on timescales of $t_\textnormal{KH}\sim t_{cc}$.

The shocked clump also will be susceptible to the RT instability, as a lighter material (the postshock flow) is accelerating a heavier one (the clump). In the plane-parallel case, one would expect RT ``bubbles'' to move into the high-density clump and RT ``spikes'' to move into the lower-density postshock flow. For a spherical clump, the curvature of the clump leads to nonaxial flow which prevents the formation of the RT spikes. The RT growth time is $t_{RT}=(g k)^{-1/2}$ \citep{chandrasekhar1961}, where $g$ is the acceleration. As discussed in e.g. \cite{fragile2004}, we may approximate the acceleration as $g\sim v_s/t_{acc}\sim v_s/(\chi^{1/2} t_{cc})$, where $t_{acc}$ is the acceleration timescale. The acceleration timescale is approximated by considering the momentum transfer of postshock material onto the clump. With this approximation, the RT growth time is

\begin{equation}\label{eq_trt}
  t_{RT} = \frac{t_{cc}}{(k r_c)^{1/2}} \quad .
\end{equation}
As with the KH instability, wavelengths of order the clump radius are seen to be the most disruptive. Therefore, as with the KH instability, for the dominant modes $t_{RT}\sim t_{cc}$. By comparing the two instabilities we see that $t_\textnormal{KH}/t_{RT}=(k r_c)^{-1/2}$. Since $t_\textnormal{KH}/t_{RT}<1$ for $\lambda<r_c$, this implies that KH instabilities at any wavenumber will grow faster and are expected to play a larger role in the destruction of the clump than RT instabilities.

The above timescales are for a hydrodynamic adiabatic shocked clump with an ideal gas equation of state. The expressions do not take into account the effects of additional physical processes which may be present in shocked clump systems. Potential important processes include radiative cooling, self-gravity, heat conduction, and magnetic fields. By careful choice of physical parameters, we may include or exclude some or all of these. Magnetic fields, for example, are known to permeate much of the ISM. The role of magnetic fields therefore may be a significant one, and magnetic fields are an important aspect addressed in many studies \citep[e.g.][]{jones1996,shin2008}. Owing to the ubiquity of shocked clumps, however, we may consider the case of nonexistent or dynamically unimportant magnetic fields without loss of generality. 

The collapse of molecular cloud cores under their own gravity is understood to be one of the progenitors of star formation. While other factors may contribute, such as the passage of a supernova shock which compresses the cloud \citep{preibisch2002} or internal energetic jets which drive turbulence within the cloud \cite{jonathan2009}, self-gravity is a universal component. The relevant timescale associated with the role of self-gravity is the free-fall timescale $t_\textnormal{ff}$,

\begin{equation}
  t_\textnormal{ff} = \left( \frac{3\pi}{32 G\rho} \right)^{1/2}
\end{equation}
As discussed in \S~\ref{numerics}, our choice of scales is such that $t_\textnormal{ff}\gg t_{cc}$, and the effects of self-gravity may be safely ignored.

As discussed by \cite{orlando2005}, thermal conduction may be ignored when the thermal conduction time scale $t_{tc}$ is either much larger or smaller than the other physical time scales. Moreover, as described in KMC94 and others, one may postulate the presence of a magnetic field which inhibits thermal conduction while remaining dynamically unimportant.

\subsection{Radiative cooling}\label{prob_cooling}
A description of the role of radiative cooling in shocks may be found in texts such as \cite{dysonwilliams1997, zeldovich2002}. Shock waves convert kinetic energy directed perpendicular to the shock front into random thermal motions. Radiative cooling removes thermal energy from the system at a rate that is characterized by some cooling law. Analytic expressions for this cooling rate typically assume a power-law form of $\Lambda(T)\propto T^{\alpha}$, where $\alpha=-1/2$ may be assumed for typical ranges of temperatures ($10^5$--$10^7$ K). From e.g. \cite{dysonwilliams1997}, the cooling occurs over a time, 

\begin{equation}
t_{cool} = \frac{3 k_B T_{ps}}{n_{ps} \Lambda(T_{ps})}\propto \frac{v_s^3}{n_a} \label{eq_tcool}
\end{equation}
where the subscript {\it ps} refers to postshock values and $n_a$ is the preshock (ambient) number density. We have made the usual assumption that $T_{ps}=p_{ps}/(\rho_{ps} k_B)\simeq (\rho_a v_s^2)/(4\rho_a k_B)$ in the strong-shock limit. The corresponding cooling length is

\begin{equation}\label{eq_lcool}
L_{cool} = v_{ps} t_{cool} = \frac34 v_s t_{cool}\propto \frac{v_s^4}{n_a}
\end{equation}
where again we use the strong-shock relation $v_{ps}=3 v_s/4$. Since the pressure remains roughly constant over the cooling length, the ideal gas equation of state dictates that the density must increase. Hence, cooling shocks allow density contrasts greater than the factor of 4 expected from the strong shock limit of jump conditions with $\gamma=5/3$ \citep{dysonwilliams1997}. 

It is often useful to consider a system in the isothermal ($\gamma=1$) regime, which may be considered a limiting case of radiative cooling. From \cite{dysonwilliams1997}, this regime occurs when $L_{cool}\rightarrow0$ and the temperature increase across the shock vanishes. In this regime $v_{ps}=v_s$ instead of $v_{ps}=3 v_s/4$. Therefore, for a fixed postshock velocity (such as a specified boundary condition), the corresponding shock velocity will be reduced in the isothermal regime compared to the adiabatic regime. We therefore expect a decrease in the shock velocity as {\it cooling strength} increases. 

There are several visual cues to the cooling strength, one of the most obvious being a greatly reduced bow shock stand-off distance for a shocked clump. We may approximate this as follows. A strong shock impinging on a dense body will form a reverse-facing bow shock whose shape is determined by the Mach number of the flow and the shape of the solid body. In the adiabatic case with a rigid ($\chi\equiv \rho_c/\rho_a=\infty$) spherical body, the distance $\Delta_0$ between the bow shock and the body is \citep{hirschel2005}

\begin{equation}\label{eq_hirschel1}
  \Delta_0 = \frac{\varepsilon r_c}{1-\varepsilon}
\end{equation}
where  
\begin{equation}\label{eq_hirschel2}
  \varepsilon \equiv \frac{\rho_1}{\rho_2} = \frac{ (\gamma-1) M_1^2+2 }{ (\gamma+1)M_1^2}
\end{equation}
with $\rho_1$, $\rho_2$ the pre- and postshock density, respectively, and $M_1$ the Mach number with respect to the preshock medium. For $M\gg 1$, $\varepsilon\rightarrow (\gamma-1)/(\gamma+1)$. Thus, for $\gamma=5/3$, $\varepsilon\rightarrow 1/4$ and the stand-off distance becomes $\Delta_0=1/3 r_c$.

What should happen to the bow shock stand-off distance when cooling is present? We approximate the effect of cooling by again considering $\gamma\sim 1$. (This approximation works well in regions of compression; it breaks down in rarefaction regions---see \cite{andy2009b}.) By inspection of Eq.~\ref{eq_hirschel1} and \ref{eq_hirschel2}, we see that in the $\gamma\rightarrow1$, $M\gg1$ limit, $\Delta_0\rightarrow0$. We find therefore that as cooling strength increases, the bow shock stand-off distance $\Delta_0$ should decrease. This rough analysis is supported by numerical simulations of cooling shocked clumps: when cooling is present, the bow shock stand-off distance decreases compared to the adiabatic case. Thus strong cooling is characterized by a bow shock wraped tightly around the clump which is, itself, undergoing strong shock processing.

We note that the situation is complicated somewhat when the rigid body is replaced by a solid body with finite density contrast $\chi$ (i.e., a clump). The bow shock reaches its steady-state stand-off distance in a time $t_{bs} = \Delta_0/v_{bs}$, where $v_{bs}$ is the bow shock velocity. As soon as the clump is hit by the shock, however, the transmitted shock begins to accelerate clump material. The bow shock may not be accelerated downstream at the same rate as the post-transmitted shock material, since the stand-off distance is determined partly by the shape of the clump, which is changing. As the stand-off distance is greater for blunt objects \citep{hirschel2005}, the axial compression of the shocked clump may push the bow shock further upstream from the clump, reducing its the bow shock's acceleration relative to the clump. Therefore, if $t_{bs}<t_{cc}$, the bow shock may achieve a stand-off distance $\Delta_{bs}>\Delta_0$. (Conversely, if $t_{bs} \gg t_{cc}$, then $\Delta_{bs}<\Delta_0$, and the bow shock may not achieve the steady state stand-off distance $\Delta_0$ given above.)

In practice, in the adiabatic case of a shocked clump with finite $\chi$, $t_{bs}<t_{cc}$, and $\Delta_{bs}$ reaches a steady state value and then remains stationary (in the preshock rest frame) for 1--2 $t_{cc}$ before being accelerated downstream. When cooling is present for the bow shock material, however, the reduced $v_{bs}$ (leading to increased $t_{bs}$) results in a bow shock which forms a small distance from the clump, and remains tightly wrapped around the clump material over the course of the destruction of the clump.

One therefore needs to define when the effects of cooling are important. In order for cooling to be dynamically significant, the cooling length should at most be $L_{cool}\sim r_c$. When $L_{cool}< r_c$, cooling occurs rapidly and can be expected to be of greater importantance in the dynamics of the flow. As in \cite{fragile2004}, we modify Eq.~\ref{eq_tcool} for the case of a shocked clump using $\chi$ and the shock jump conditions as, 

\begin{equation}\label{eq_tcoolclump}
t_{cool} = \beta \frac{v_s^3}{\chi^{3/2} n_c} = 1.1\times10^2 \frac{ M^3\ T_a^{3/2} }{ \chi^{5/2}\ n_a }\ s
\end{equation}
where the subscript $a$ refers to the ambient (preshock) material, and $\beta=6.61\times 10^{-11}\ cm^{-6} s^4$.

Similarly we write a cooling length for the clump material as
\begin{align}
  L_{cool} &= v_{ps,c} t_{cool} \simeq \frac{v_s}{\sqrt{\chi}} t_{cool} \nonumber \\
        &= \beta \frac{v_s^4}{\chi^2 n_c}= 1.3\times10^6 \frac{M^4 T_a^2}{\chi^3 n_a}\ cm \label{eq_lcoolclump}
\end{align}
where $v_{ps,c}$ is the postshock velocity in the clump. 

From a numerical standpoint, adequate resolution of the cooling length is important to ensure accurate capturing of the energy loss as the postshock fluid parcels cool. Inadequate resolution may lead to an underestimation of the cooling rate -- or, equivalently, an overestimation of the cooling length. In particular, the contact discontinuity between the upstream-facing clump bow shock and the downstream-facing transmitted shock must be adequately resolved in order to properly allow the growth of instabilities in the ablation region. This is discussed further in \S\S~\ref{analysis} and \ref{discussion}.
\section{Numerics}\label{numerics}
\subsection{The AstroBEAR code and the computational domain}\label{num_astrobear}
For the simulations we employed the AstroBEAR code using a 2D computational grid including the effects of axisymmetry ({\it 2.5D}) as an operator-split source term. Employing a 2D grid with such a source term is widely employed in situations where variation in the polar angle $\phi$ is expected to be minimal. Such is the case here with an initially planar shock propagating in $z$ and a spherical clump located on the axis. We may therefore utilize a 2.5D simulation to approximate the results of a fully 3D simulation at reduced computational cost. However, to support our conclusions from the 2.5D data, we have undertaken a supplementary 3D simulation. This is discussed in \S~\ref{results}.

The AstroBEAR code is a parallel AMR Eulerian hydrodynamics code with capabilities for MHD in two- and three-dimensions. There are several schemes of varying order available for the user; here the code solves the fluid equations using a second-order MUSCL scheme and treats cooling and cylindrical symmetry as Strang-split stiff source terms. AMR is employed to more accurately track regions of interest, determined by gradients in fluid variables. Details on AstroBEAR may be found in \cite{andy2009}. The inclusion of simple multiphysics is possible, tracking helium and hydrogen ionization; here hydrogen is assumed. The cooling curve of \cite{dalgarno1972} is employed; refinements on this method are being undertaken. Advected tracers are used to track the preshock, postshock, and clump material throughout the simulation. 

The computational domain has extent in $z$ and $r$ of 8$r_c$ and 2$r_c$, respectively. Our computational domain is the same as that in \cite{fragile2004} and comparable to the 3D domain of \cite{orlando2005}. The physical parameters are listed in Table~\ref{tab_parameters}. The clump radius $r_c$ is 50 AU---however, this length scale is not restrictive as our results scale to larger or smaller cloud sizes via appropriate choice of dimensionless numbers; see below. The clump is located on the axis of symmetry at $z=2 r_c$ and has a density contrast with the ambient (preshock) medium of $\chi\equiv\rho_c/\rho_a=10^2$. The preshock temperature is $T_a=2000\ K$. A shock of mach number $M=50$ was placed initially at $z=0.5 r_c$, with postshock conditions imposed on the left ($z=0$) boundary. The top and right boundaries have 0th-order extrapolating boundary conditions. The initially imposed shock crosses the domain in $\sim1.1 t_{cc}$. The time for all the clump material to blow off the grid is longer: the simulations all continue until all clump material has left the grid which occurs at approximately $3 t_{cc}$.

\begin{table}[htbp]\centering\small
\begin{tabular}{lrr}
\hline\hline
Quantity          & Ambient           & Clump \\
\hline
Density           &  $10^2\ cm^{-3}$  & $10^4 cm^{-3}$\\
                  &                   & $\chi=10^2$  \\
Temperature       & 2000 K            & 20 K \\
Sound speed       & 5.2 $km\ s^{-1}$  & 0.52 $km\ s^{-1}$ \\
Shock velocity    & $260\ km\ s^{-1}$ & $26\ km\ s^{-1}$ \\
Shock Mach number & 50                & 50 \\
Radius            &                   & 50 AU \\
$t_{cool}$ (M=50) & 3.6 yr            & 0.036 yr \\
$t_{cool}/t_{cc}$ & 0.40              & 0.004 \\
$L_{cool}$        &                   & $1.8$ AU (0.04 $r_c$) \\
$t_\textnormal{ff}$&                  & $5\times10^5$ yr \\
\hline
\end{tabular}
\caption{Physical parameters of the simulations for the ambient (preshock) and clump material. The shock velocities are for the {\it external} shock and transmitted ({\it internal}) shock, respectively.\label{tab_parameters}}
\end{table}

The resolutions for the 8 runs are given in Table~\ref{tab_resolutions}. AMR was employed for the higher resolution runs, up to a maximum of 2 additional levels, where each subsequent level doubles the resolution. The ``effective resolution'' listed is the equivalent fixed-grid resolution for the highest refined level. Our lowest-resolution run had N=12 cells$/r_c$. Each run thereafter doubled the resolution, to N=24, 48, 96, 192, 384, 768, and 1536 cells$/r_c$. We refer to each run according to these values as $R_N$.

\begin{table}[htbp]\centering\small
\begin{tabular}{rrrr}
\hline\hline
Run & Effective Resolution & $\Delta x/L_{cool}$& $\Delta x/L_{cool,inspec.}$\\
\hline
$cells/r_c$ & $cells\ in\ z\times r$  & & $Approx.$\\
12  &    $ 96\times    24$  &   -- & --       \\
24  &    $192\times    48$  &    1 & $\sim 1$ \\
48  &    $384\times    96$  &    2 & 1-2      \\
96  &    $768\times   192$  &    4 &  3       \\
192 &  $1,536\times   384$  &    7 &  5       \\
384 &  $3,072\times   768$  &   15 &  7       \\
768 &  $6,144\times 1,536$  &   29 & 11       \\
1536& $12,288\times 3,072$  &   58 & 14       \\
\hline
\end{tabular}
\caption{Resolutions of the 8 simulations, from 24 to 1,536 cells per clump radius $r_c$. The third column gives the number of cells per cooling length $L_{cool}$ from Eq.~\ref{eq_lcoolclump}, and the fourth column via visual inspection (see \S~\ref{analysis}). At \rr{12} the cooling length was unresolved.\label{tab_resolutions}}
\end{table}

We note that the specific scales we have chosen are not critical for our results, with the exception of the preshock temperature, as the cooling rate is a function of temperature. The scales of importance are the dimensionless ratios such as the cooling length versus the clump radius $L_{cool}/r_{c}$, the cooling time versus the cloud-crushing time $t_{cool}/t_{cc}$, and the density contrat $\chi$. We expect the results presented here to be demonstrative of any environment with similar dimensionless numbers. For comparison, for our choice of parameters, $t_{cool}/t_{cc}=0.004$, identical to case E5 of \cite{fragile2004} which used a cloud of radius $R_c \sim 100 pc$. 

The cooling length for the postshock clump material (see \S~\ref{problem}) is $L_{cool}<0.1 r_c$, indicating that cooling is important in the clump. Since the preshock temperature is $T_a=2000\ K$, cooling also will be important for shocks in the ambient material. As we will see, the choice of temperatures results in the strongest cooling occuring in the clump bow shock.

We choose a sharp cloud boundary for our simulations, such that at radius $r<r_c$ the clump has density $\rho_c$ and temperature $T_c$, and at radius $r>r_c$ these values are the preshock density and temperature. This yields a step-function profile. The choice of clump profile is a matter of some interest. Most studies have adopted a sharp boundary\citep[e.g. KMC94,][and others]{fragile2004, orlando2005, mellema2002}, while some recent studies have chosen to enclose a dense ``core'' within a ``cloud'' of larger radius which has some decreasing profile in density and corresponding transition in temperature to the preshock ambient medium \citep{poludnenko2002, patnaude2005, nakamura2006}.  As seen in \cite{nakamura2006}, the chief dynamic result of this is to soften the impact of the shock and increase the growth time of instabilities. We adopt a sharp boundary for conceptual simplicity. Such a boundary imposes a grid-based set of perturbations on the clump interface which in principle will influence the growth of Kelvin-Helmholtz (KH) and Rayleigh-Taylor (RT) instabilities. However, these perturbations have wavelengths $\lambda\ll r_c$. Under the assumption of \S~\ref{problem} that the dominant modes are those with $\lambda\sim r_c$, these perturbations therefore should have minimal influence on how the solution varies by resolution. This assumption appears borne out by the results of our 3D simulation, discussed in \S~\ref{results}, which feature little or no grid-based artifacts.

\section{Results}\label{results}
\begin{figure}[htbp]
\plotone{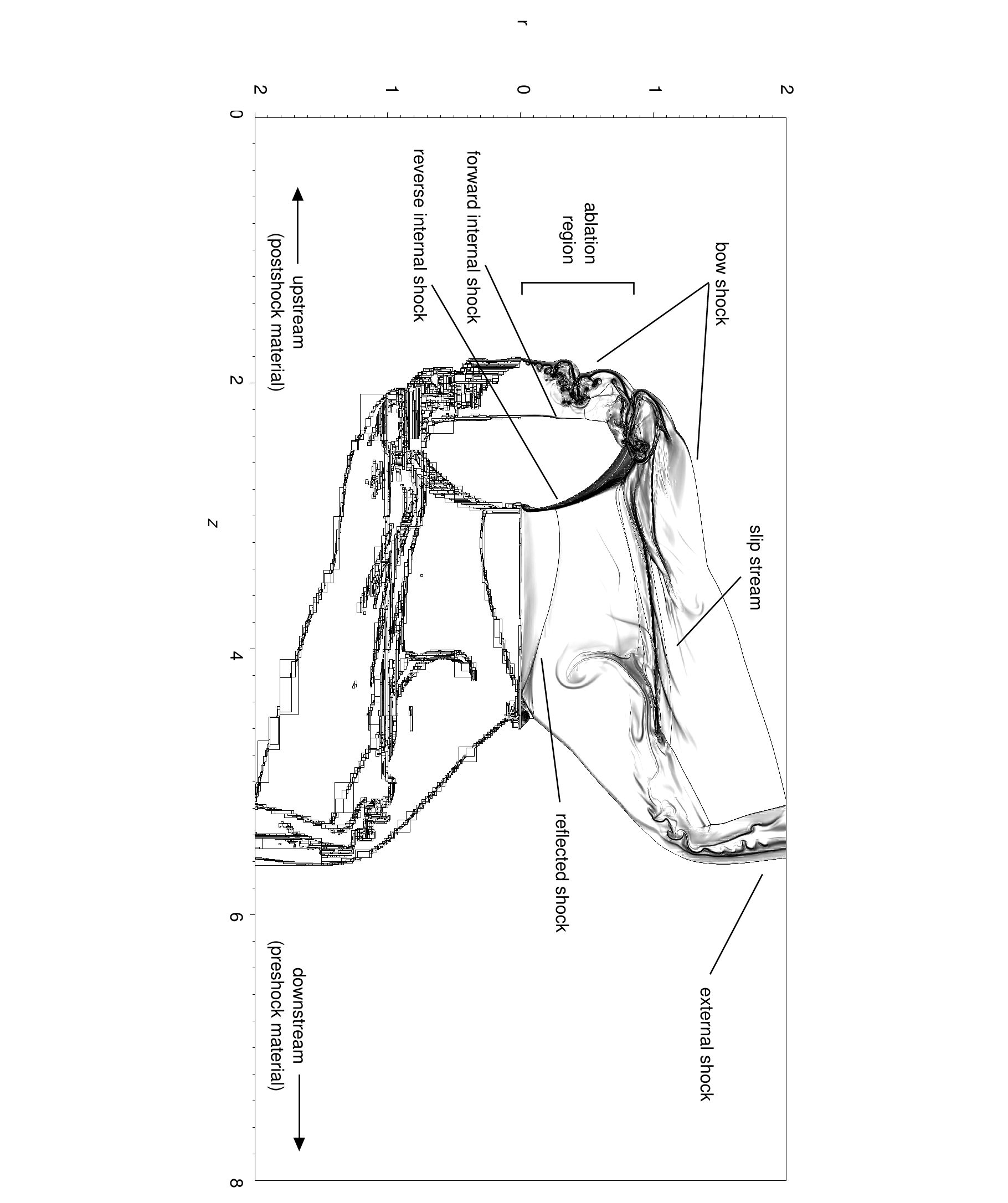}
\caption{{\small Important features of the flow are given in a synthetic Schlieren image of run \rr{1536} at $t\sim 0.5 t_{cc}$. The image has been reflected about the axis of symmetry, with the reflection showing the location of AMR refined regions. The bow shock wraps tightly around the clump from the strong cooling. The clump surface is ablated by its interaction with the postshock flow. A slip stream forms behind the clump. Transmitted shocks propagate internally through the clump. The external shock is susceptible to the cooling instability, and a conical reflected shock forms off of the axis of symmetry which reengages the flow.}}
\label{fig_diagram}
\end{figure}

The same basic physics applies for all runs; see Fig.~\ref{fig_diagram}. A Mach 50 shock moving in the positive $z$ direction with velocity $v_s$ impacts the clump located on the axis at $z=2 r_c$ with density contrast $\chi=100$. This impact creates an internal (transmitted) shock which propagates through the clump at a slower velocity $v_{s,c}\sim v_s \chi^{-1/2}$. Both the internal and external shocks are radiatively cooling.  {\it Downstream} (positive $z$) is rightward on the page; {\it upstream} is to the left.

In a time $t\sim t_{cc}$ the external shock propagates off the edge of the computational domain (located at $z=8 r_c$). A reverse-facing bow shock forms which has a small stand-off distance from the edge of the clump (see \S~\ref{prob_cooling}). This implies that the bow shock/contact discontinuity region is resolved by a correspondingly small number of cells. This important point is discussed further in \S~\ref{analysis}. Note as the internal shock propagates through the clump, material is stripped off the leading edge of the clump by KH instabilities and the bulk post-shock flow. We refer to this region (between the bowshock and the contact discontinuity) as the {\it ablation front}. RT bubbles propagate into the clump, while RT fingers are destroyed by the postshock flow. Finally, the downstream flow exhibits multiple interacting shocks.

\begin{figure}[htbp]
\centering
\plotone{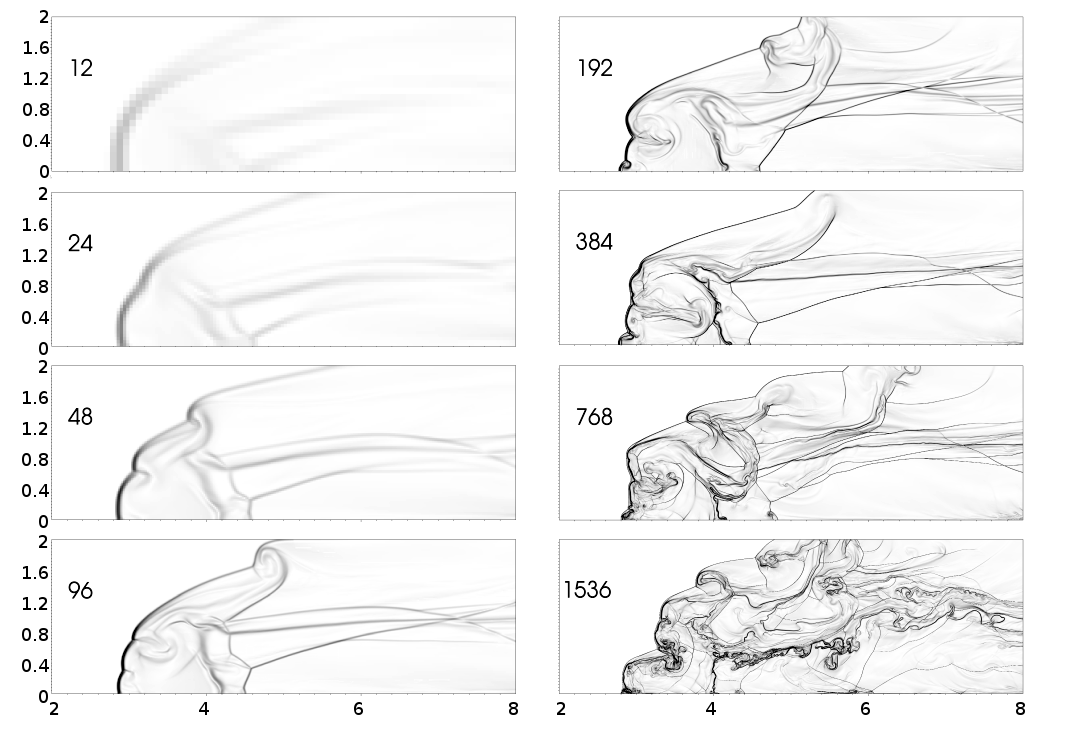}
\caption{{\small Synthetic Schlieren images of the 8 simulations of radiatively cooling shocked clumps in 2.5D are shown at time $t=1.44 t_{cc}$. The Mach number of the shock is $M=50$. Spatial units are in clump radii $r_c$. A clump was located initially at $z=2 r_c$ and has a density contrast with the preshock ambient medium of $\chi=100$. The computational domain in $z$ and $r$ is $8 r_c$ by $2 r_c$,respectively -- in the images the computational domain has been truncated on the left edge at $z=r_c$. Each panel is labelled with the initial number of cells per clump radius (see \S~\ref{numerics} and Table~\ref{tab_resolutions}). Visually, the simulations appear to show self-convergence up to a point (384 or 768 cells/$r_c$), after which they do not. The parameter regime is such that both the external and transmitted shocks are cooling (hence the extremely small clump bow shock stand-off distance, see \S~\ref{problem}). See \S~\ref{results} for more details. }}
\label{fig_schlieren}
\end{figure}

We now visually compare the eight different resolutions. Figure~\ref{fig_schlieren} gives synthetic Schlieren images of the simulations. Synthetic Schlieren images are a useful tool for investigating dynamic systems of shocks because they highlight the locations of gradients in density. Measured in cells per clump radius, the resolutions are 12, 24, 48, 96, 192, 384, 768, and 1536. Each panel in the figure is a different resolution, and all panels are at the same time in the simulation, $t=1.44 t_{cc}$. (Note that in the image the left border of each panel is truncated at $z=r_c$.) At this point in the simulation, the internal shock has completely crossed the clump.

Are the simulations converging? In a typical demonstration of self-convergence, authors will show qualitatively and quantitatively that as resolution increases, differences between successive resolutions decrease. We discuss quantitative apects the simulations and their convergence in \S~\ref{analysis}. Qualitatively, with increasing resolution there are visual cues of convergence, namely the presence of structures whose appearance remain basically the same, but ``sharpen'' with higher resolution. The ablation front remains at roughly the same location in $z$, while the width (extent in $z$) of the feature decreases with resolution. There is an extended feature starting at the rear of the clump ($z=3.0$, $r=1.0$) which extends downstream off the edge of the grid. This feature is a result of a conically shaped shock reflecting off the symmetry axis and reengaging material flowing around the clump. This results in a multiple-shock, axially-aligned structure. In cases through \rr{768}, this structure is recognizable and is sharpened similar to the ablation front. At \rr{1536}, this structure is destroyed apparently due to enhanced vorticity. Indeed, \rr{1536} appears to show the most disparate structures of any of the cases. This suggests that, while the simulations appear to converge up to a point, above a certain resolution key features of the simulations diverge as new physically relevent length scales are captured.

\begin{figure}[htbp]
\centering
\plotone{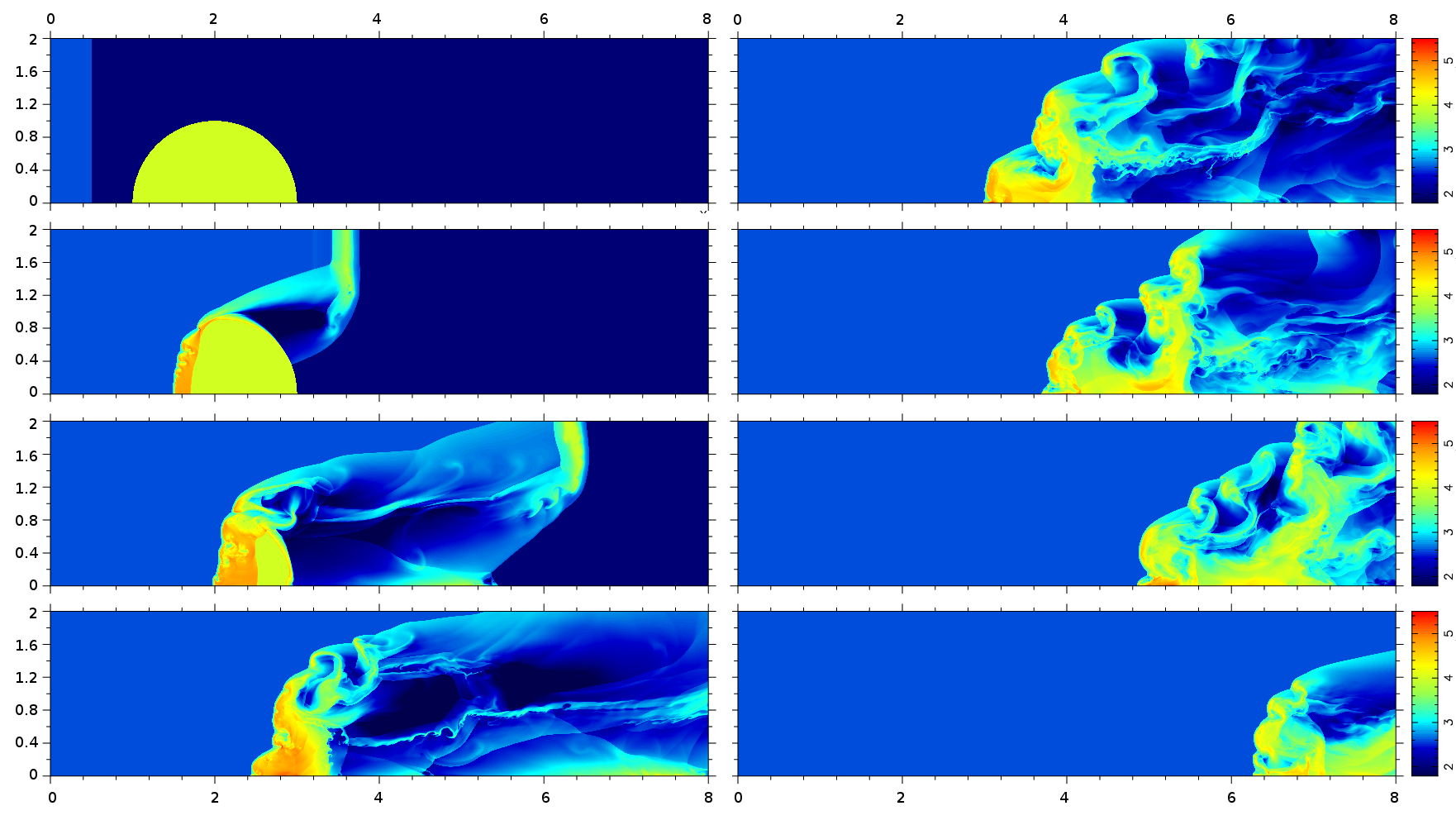}
\caption{{\small Eight panels $a$--$h$ show false-color representations of the logarithm of density for run \rr{1536} at eight times in the simulation, at $t$=-0.04, 0.4, 0.8, 1.2, 1.6, 2.0, 2.4, and 2.8 $t_{cc}$, respectively. Both external and internal shocks are radiatively cooling. The clump bow shock is wrapped tightly around the clump. All of the clump material is in motion by the fourth panel, and RT and KH instabilities destroy the clump after a few $t_{cc}$. See \S~\ref{results} for further details.}}
\label{fig_1536}
\end{figure}

How important is this apparent disparity of \rr{1536}? In other words, how does the evolution of case \rr{1536} compare with a case having resolution which in the adiabatic regime would be considered ``sufficient''? We address this question by examining the time evolution of case \rr{1536}, and comparing this evolution with case \rr{192}, which is at a resolution that would be considered sufficient for an adiabatic simulation. Figure \ref{fig_1536} follows the time evolution of the highest resolution run, \rr{1536}, with a false color representation of logarithmic density. The eight panels $a$--$h$ depict the simulation from $t$=-0.04 to 2.8 $t_{cc}$. Panel $a$ of Fig.~\ref{fig_1536} shows the initial conditions for the highest-resolution simulation. In panel $b$, at $t=0.40 t_{cc}$, the external shock has reached $z=4 r_c$. The shock is not a sharp discontinuity owing to cooling. The internal shock has propagated $\sim2/3 r_c$ into the clump. An upstream-facing bow shock has formed around the clump, and is wrapped tightly around the clump surface. The surface of the bow shock at distances greater than $r_c$ from the cloud center is smooth. The ablation surface of the clump is being disrupted by RT and KH instabilities. The amplitude of the RT bubbles increases with radius, suggesting that there is a KH component which enhances the growth: further from the axis, the post-shock flow is increasingly oblique to the clump surface, implying that the relative velocity is higher and so from Eq.~\ref{eq_tkh} the KH growth rate is increased. It is worth noting that there is a clear separation between the internal shock at $z=1.7$ and the ablation front at $z=1.5$. This is in contrast to other studies where the cooling rate is so high that the distance between these features essentially goes to zero \citep[e.g.][]{mellema2002,fragile2004}. This important point is discussed further in \S\S \ref{problem}~\&~\ref{analysis}. We also note the presence of a high-density peak at roughly $z=1.0$, $r=0.8$, whose evolution will be followed over the next few panels for comparison with \rr{192}.

In panel $c$ at $t=0.8 t_{cc}$, the external shock has reached $z=6.5$ and continues to be disrupted. The internal shock has propagated about 75\% of the way through the clump. The portion of the external shock seen wrapping around the clump in panel $b$ has by this point reflected off the axis of symmetry, and the accompanying pressure increase has formed an upstream-propagating internal shock which has propagated into the rear clump by about 20\%. The high-density peak noted in panel $b$ has become separated from the clump. Though not readily apparent in this figure, its separation is caused by a fast-growing RT instability which forms at a radius just inside that of the peak, which then quickly propagates through the clump. The RT bubble is evident in the small downstream-facing shock emerging from the clump at roughly $z=3.0,\ r=0.7$. Overall, the downstream-facing internal shock front remains smooth in directions perpendicular to the flow, though its shape is distorted (especially at larger radii) by the instabilities and other transmited shocks compressing the clump. The front of the clump continues to be processed by the post-shock flow via the KH and RT instabilities.

Panel $d$ depicts $t=1.2 t_{cc}$. By this point the external shock has propagated off the grid. The region downstream of the clump exhibits KH instabilities, as well as the presence of numerous shocks. The shocks are a result of reflections off the axis as well as episodes of shock generation upstream near the clump. (The small shock noted in the previous panel is an example of one such episode.) There are also two large rarefaction regions downstream of the clump. The high-density peak noted in panel $b$, which separated from the clump in panel $c$, has further fragmented into two distinct vortical features in this panel, located at $(z,\ r)=$ $(3.2,\ 1.4)$ and $(3.6,\ 1.7)$, respectively. The downstream-facing internal shock has propagated entirely through the clump and has collided with the upstream-propagating internal shocks. The shape of the clump ablation front is noticeably different than in earlier panels. The instabilities at the front have allowed a large portion of clump material to shear radially, leaving only material close to the axis behind. All of the clump material is in motion.

Panels $e$ through $h$, at times $t=1.6,\ 2.0,\ 2.4, \&\ 2.8\ t_{cc}$, respectively, exhibit the continued processing of the clump by the postshock flow. Clump material is episodically, rather than continually, stripped from the clump. The shape of the bow shock changes during these episodes as it remains tightly wrapped around the clump material. 

\begin{figure}[htbp]
\centering
\plotone{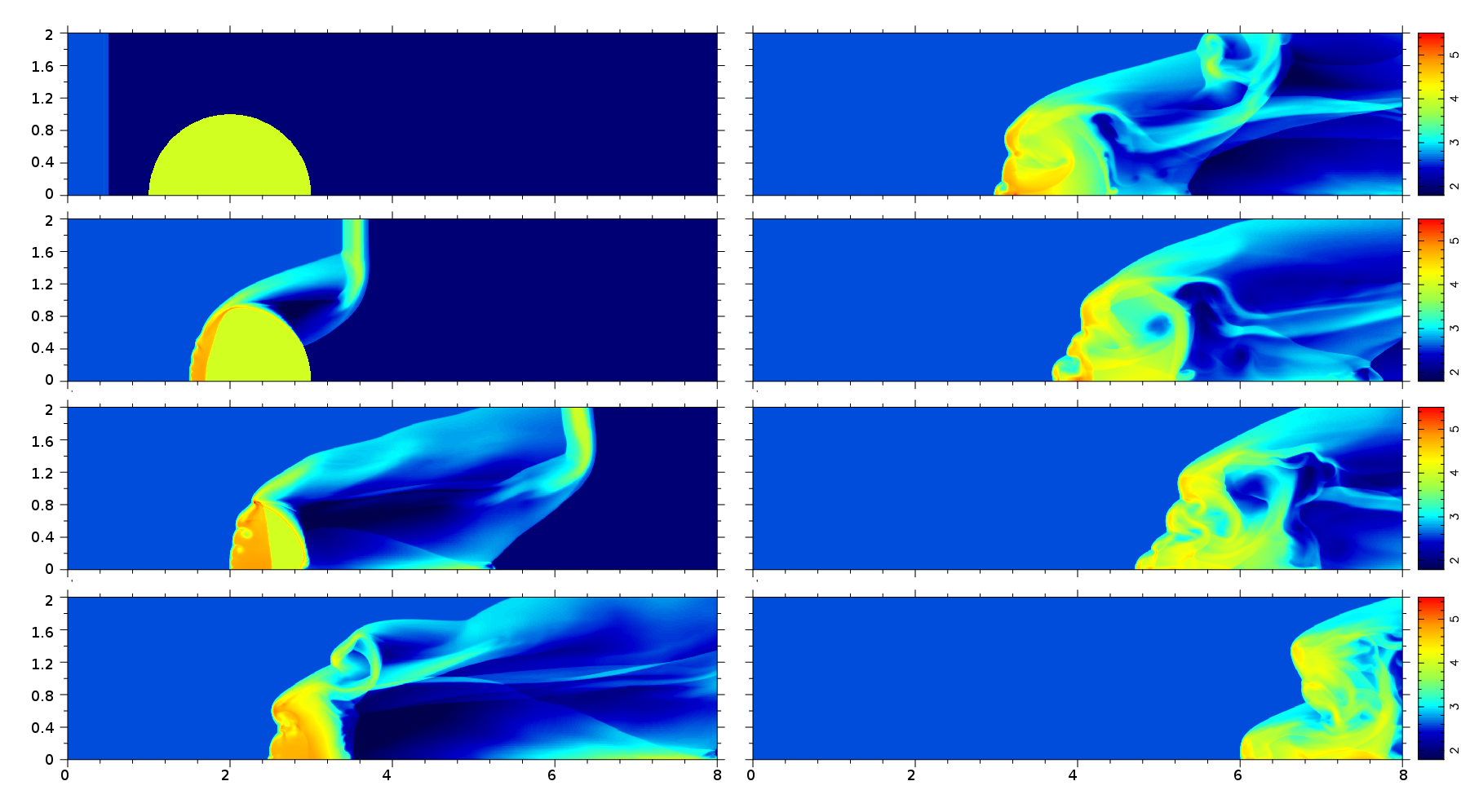}
\caption{{\small Eight panels show false-color representations of the logarithm of density for run \rr{192} at eight times in the simulation, at $t$=-0.04, 0.4, 0.8, 1.2, 1.6, 2.0, 2.4, and 2.8 $t_{cc}$, respectively. The figure may be compred with Fig.~\ref{fig_1536}. While overall the evolution is similar there are notable differences, including the shape of the internal shock and the lack of vortices in the downstream region.}}
\label{fig_192}
\end{figure}

We now consider the evolution of case \rr{192}, in comparison to that of \rr{1536}. Figure \ref{fig_192} shows eight images of case \rr{192} at the same times as shown in Figure \ref{fig_1536}. Overall, the behavior is similar. In particular, the axial location of the ablation front is consistent with that seen in Fig.~\ref{fig_1536}. The high-density clump parcel noted in Fig.~\ref{fig_1536} is present and evolves similarly, appearing in panel $c$ at $z=2.4$, $r=0.8$. However, in this case it remains a single vortex rather than multiple vortices. Also in panel $c$, the shape of the internal shock is different, as are the RT bubbles behind the clump ablation front. Here, the shock front remains smooth and unaffected by the RT bubbles up to the time when the shock crosses the clump, whereas in Fig.~\ref{fig_1536} the instabilities propagated up to the shock front, causing it to distort from planar. Overall, fewer vortices are noted, especially in the downstream region. We conclude that, visually, the behavior is unclear as to convergence, and deserves more quantitative investigation (\S~5).

\begin{figure}[htbp]
\centering
\plotone{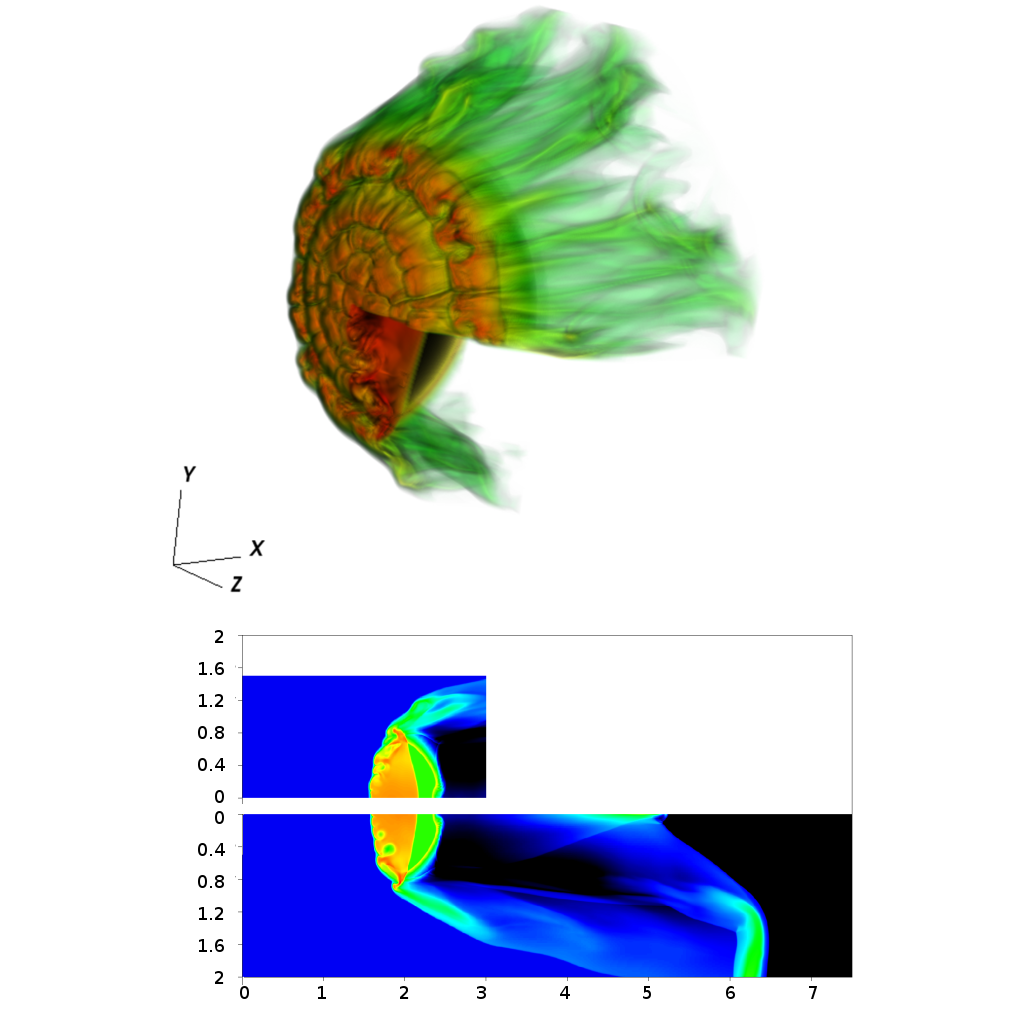}
\caption{{\small {\it Top panel}: A volumetric representation of logarithm of density for run \rr{192,3D} at $t=0.80 t_{cc}$. The simulated region occupied one quadrant of space only; it has been reflected to two other quadrants in the image. The location of the transmitted shock is evident, as are the RT and KH instabilities at the clump ablation surface. Streams of material stripped off the clump surface are evident as well. {\it Bottom panel:} A comparison between runs \rr{192} and \rr{192,3D}. A slice through the 3D domain of log density is shown in comparison to the 2.5D result at the same time, $t=0.80 t_{cc}$, corresponding to panel $c$ of Fig.~\ref{fig_192}. The results are in good agreement, differing slightly in the growth of RT bubbles into the cloud and the shape of the rollover region at the top of the clump. This suggests good reliability of the rest of the simulations presented.}}
\label{fig_3d}
\end{figure}

We undertook a single cooling 3D run for comparison with the 2.5D results, \rr{192,3D}, depicted in Fig.~\ref{fig_3d}. Both panels in the figure are at the same time, ($t=0.8 t_{cc}$). The top panel depicts the logarithm of density in a volumetric plot. The simulation followed the $\pm$x/+y/+z quadrant of the clump sphere, which has been reflected around to the -y/+z and -y/-z quadrants in the figure. The internal shock is evident in the image, as are the instabilities propagating into the clump. Note however that the 3D nature of the simulations allows us to capture non-axisymetric flow patterns, and the cloud surface is seen to be ablating in a nonuniform manner with respect to $\phi$. Streams of material are evident flowing downstream from non-axisymetric clumps on the cloud surface.  Information about the surface of the clump is propagated downstream as increased or decreased axial flow downstream. These non-axisymetric modes may have been seeded due to grid-based effects from the sharp cloud boundary.  It is clear however that by this time in the evolution their effects on the cloud surface are minimal. If grid-based effects played a large role in the evolution one would expect corresponding features on the order of $\Delta x$. In fact, all visible perturbations at this time are appreciably larger, suggesting that the sharp cloud boundary does not play a significant role after providing an initial perturbation seed.

The bottom panel depicts a ``slice'' through the computational domain in the y=0 plane of \rr{192,3D} compared directly to \rr{192} at the same time, which gives the same resolution but in 2.5D. A high degree of agreement is seen. The shapes and locations of the internal shock are nearly identical. The ablation fronts are very similar, though the 2.5D case has a more-developed RT bubble at $z=2.0$, $r=0.6$ than the 3D case. The rear of the clump, and the other transmitted shocks, also are quite similar. The shapes of the bow shock at radii larger than $r_c$ differ slightly, with the 2.5D case having a smoother rollover than the 3D case. Finally, the conical reflected shock is more prominent in the 2.5D case than in 3D. This good agreement suggests that the 2.5D simulations are reasonably reliable.
\section{Analysis}\label{analysis}
\subsection{Convergence of global quantities}\label{an_convergence}
We wish to evaluate quantitatively how the simulations change with resolution; namely, whether they converge.  We begin by examining the behavior of global quantities related to the clumps. A typical characterization method of shocked clump simulations is to integrate certain physical quantities over the computational domain, and witness how these quantities vary with time \citep[e.g.][]{nakamura2006, pittard2009, shin2008}. KMC94 employed a two-fluid model, with one fluid being the clump and the other fluid the pre- and postshock medium, allowing them to explicitly track the clump fluid parcels. \cite{nakamura2006} employed a single-fluid model and used a density cutoff to prescribe ``clump'' material. Several authors track the quantities by including color (advected) tracers to differentiate the clump from the ambient environment \citep[e.g.][]{pittard2009, shin2008}. Our advected tracers allows us to repeat this characterization.

We define six global quantities as in \cite{nakamura2006}: clump mass, average axial velocity, radius in the cylindrical direction, radius in the axial direction, and velocity dispersions in the cylindrical and axial directions, respectively.

The clump mass is defined as 

\begin{equation}\label{eq_clumpmass}
m_c=\int_{C>C_{min}}\rho\ dV
\end{equation}
where the integration is performed only over cells which have a fraction of clump material above a cutoff $C>C_{min}$, $0\le C \le 1$. For the results presented in \S~\ref{analysis} we use $C_{min}=0.1$---however we find the behavior of most quantities robust to the specific choice of $C_{min}$ with the obvious exception of $m_c$.

The radial and axial radii of the clump are defined as

\begin{equation}\label{eq_radrad}
a=\left(\frac52 \langle r^2 \rangle \right)^{1/2} \ ,
\end{equation}
\begin{equation}\label{eq_radax}
c=\left[5 \left(\langle z^2 \rangle - \langle z\rangle^2 \right)\right]^{1/2} \ ,
\end{equation}
and the radial and axial velocity dispersions are similarly defined as
\begin{equation}\label{rq_disprad}
\delta v_r = \langle v_r^2 \rangle^{1/2} \ ,
\end{equation}
\begin{equation}\label{eq_dispax}
\delta v_z = \left( \langle v_z^2\rangle - \langle v_z\rangle^2\right)^{1/2} \ ,
\end{equation}
where for a quantity $q^n$,
\[
\langle q^n \rangle \equiv m_c^{-1} \int_{C>C_{min}} \rho q^n dV \ .
\]
Finally, the average axial velocity is $\langle v_z \rangle$.

To quantitatively address convergence, we use the standard definition of relative error,

\begin{equation}
  \varepsilon(M,N) = \frac{|Q_N-Q_M|}{Q_N}
\end{equation}
where $Q_N$ ($Q_M$) is a global quantity defined as above for run \rr{N} (\rr{M}), and $N>M$.

Convergence implies that, given a resolution \rr{N}, 

\begin{equation}
\varepsilon(N,2N) < \varepsilon(N/2,N) \quad .
\end{equation}
That is, if simulations with higher resolutions converge on the solution, the errors between higher and higher resolutions should decrease. This implies that a progression of relative errors, showing monotonic decrease, is necessary to determine convergence. Conversely, a lack of monotonic decrease in $\varepsilon(M,N)$ may suggest at least two possibilities. One possibility is that the simulations may be of such inadequate resolution that an increase in resolution from \rr{N} to \rr{2N} returns a solution functionally as poor as that of \rr{N}.  On the other hand, it may be simply that the simulations are not converging. Such would be the case, for example, in a highly nonlinear or fully turbulent simulation, where even at a given resolution, minor changes to the initial conditions would yield differing evolution even if statistical measures were steady. A third possibility, that ``plateaus'' of convergence may exist, is discussed \S~\ref{discussion}.

We may therefore ask how the relative error $\varepsilon(M,N)$ varies over our set of simulations, and whether through a progression of increasing resolution from \rr{M} to \rr{2M} results in a monotonic decrease of $\varepsilon(M,2M)$. Table~\ref{tab_relerrors} lists the relative errors between all pairs of runs for one global quantity, clump mass, as defined above, at time $t\sim t_{cc}$. At this point in the simulation, the external shock has completely passed the clump, and the transmitted shock has propagated roughly halfway through the clump. The columns in Table~\ref{tab_relerrors} represent runs at resolutions $N$ and the rows represent runs at resolutions $M$. The leading diagonal, then, shows the $\varepsilon(M,2M)$ pairs of interest. Note that this is a representative time fairly early in the simulation, $t\sim t_{cc}$; at later times the values differ---see \S~\ref{an_convtime}.

The values on the diagonal do not monotonically decrease, meaning that increases in resolution do not consistently reduce the relative error. Note moreover that a single value of $\varepsilon(M,N)$ is not a sufficient condition to determine convergence. For example, $\varepsilon(12,24)=0.06$, but $\varepsilon(12,1536)=0.20$. If only simulations at \rr{12} and \rr{24} were performed, one might conclude based on $\varepsilon(12,24)$ that \rr{12} was converged, when in fact they both differ largely from \rr{1536}. Both \rr{12} and \rr{24} are so insufficiently resolved compared to \rr{1536} that $\varepsilon(12,24)$ means little on its own. (Of course, the derived value of $L_{cool}$ would be an independent indicator that simulations at this resolution were not sufficiently resolved.)

This illustrates why a progression of $\varepsilon(M,N)$ is necessary. In fact, for convergence to occur both $\varepsilon(N,2N)$ and $\varepsilon(N,N_{max})$ must be motonically decreasing and small (where by $N_{max}$ we mean the simulation of maximum resolution). A small $\varepsilon(N,N_{max})$ ensures the goodness (accuracy) of the solution at resolution \rr{N}. A small $\varepsilon(N,2N)$ ensures that the goodness is real and not due to a selection effect: since the temporal evolution of the global quantities differs by resolution (as we will see below), it may be possible to choose an instant in time at which two resolutions seem to agree, misrepresentative of the general behavior over time. It is unlikely, however, that at such a time both $\varepsilon(N,N_{max})$ and $\varepsilon(N,2N)$ will be small, unless the simulation does in fact converge.

In Table~\ref{tab_relerrors} we satisfy the criteria that $\varepsilon(N,N_{max})$ decreases with increasing $N$. However, these data are from early in the simulation, $t\sim t_{cc}$, and with the most well-behaved global quantity, $m_c$. We do not see the same behavior with other quantities, and we see no convergence later in time, as we now discuss.

\begin{table}[htbp]\centering\small
\begin{tabular}{r|rrrrrrrr}
\hline\hline
    &   24  &   48  &  96  & 192  & 384  & 768  & 1536 \\
\hline
12  & 0.063 & 0.013 & 0.17 & 0.18 & 0.23 & 0.22 & 0.20 \\
24  &       & 0.072 & 0.11 & 0.12 & 0.17 & 0.16 & 0.15 \\
48  &       &       & 0.04 & 0.06 & 0.11 & 0.10 & 0.08 \\
96  &       &       &      & 0.01 & 0.07 & 0.06 & 0.04 \\
192 &       &       &      &      & 0.06 & 0.05 & 0.03 \\
384 &       &       &      &      &      & 0.01 & 0.03 \\
768 &       &       &      &      &      &      & 0.02 \\
\hline
\end{tabular}
\caption{{\small Comparison of relative errors for each pair of simulations. The relative error is $\varepsilon(M,N)=|Q_N-Q_M|/Q_N$, where $Q_N$ is a global quantity for run \rr{N} and $N>M$. The quantity here is measured clump mass at $t\sim t_{cc}$. The columns are runs \rr{N} and the rows runs \rr{M}. If convergence were being witnessed, the values along the leading diagonal ( $\varepsilon(M,2M)$ ) would be monotonically decreasing. In fact, a large increase in relative error is noted at $\varepsilon(192,384)$, roughly the resolution at which the resolution typically is assumed to be sufficient. See \S~\ref{analysis} for details.\label{tab_relerrors}}}
\end{table}
\subsection{The behavior of global quantities over time}\label{an_convtime}

\begin{figure}[htbp]
\centering
\plotone{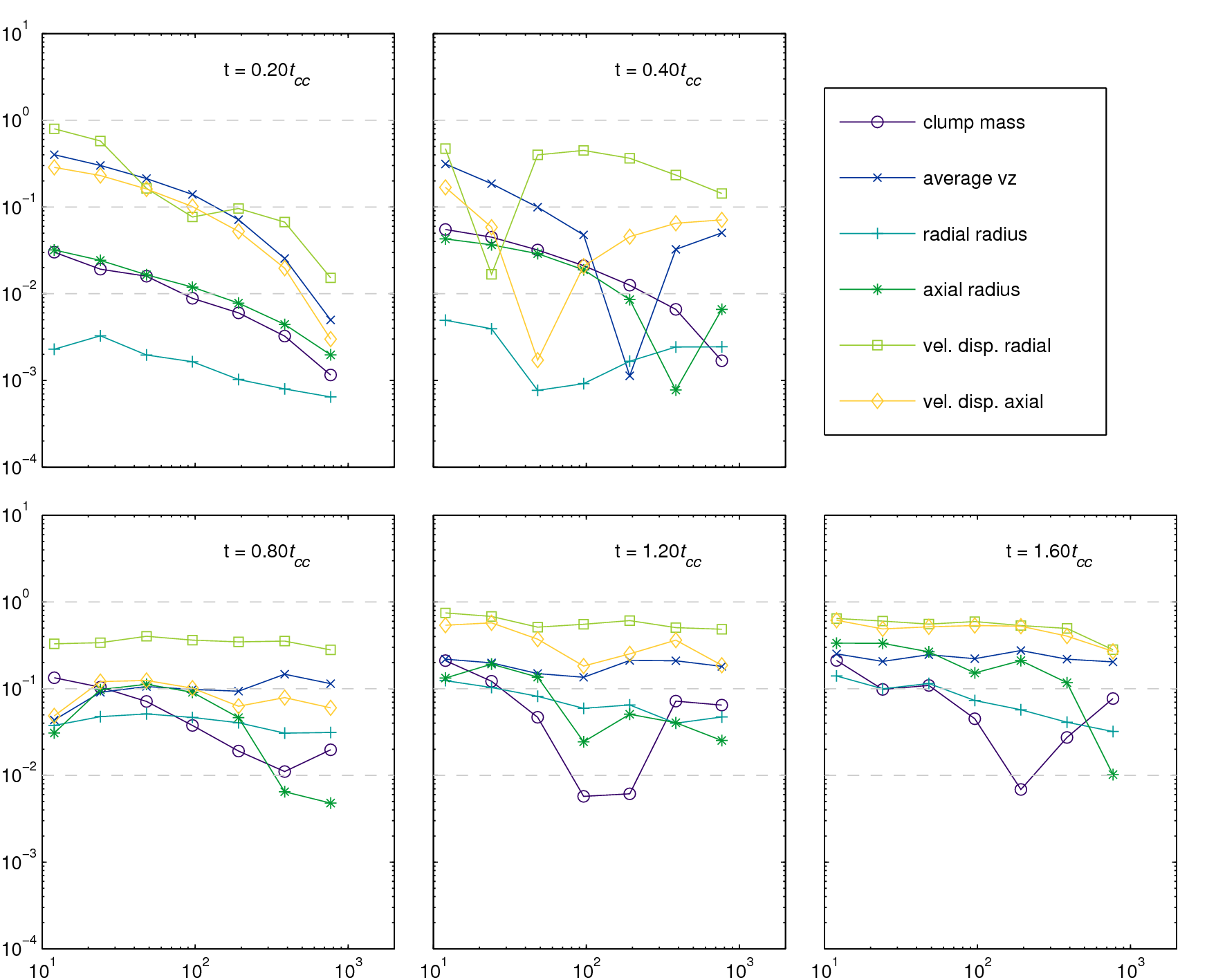}
\caption{{\small Plots showing relative error $\varepsilon(N,1536)$ as defined in \S~\ref{an_convergence} are presented at five times in the simulation, $t=0.20 t_{cc}$, $t=0.40 t_{cc}$, $t=0.80 t_{cc}$, $t=1.2 t_{cc}$, and $t=1.6 t_{cc}$. Only very early in time does convergence appear; subsequent plots show little consistent indication of convergence. See \S~\ref{an_convtime}.}}
\label{fig_eps_at_4}
\end{figure}

It is common to show convergence results for $\varepsilon(N,N_{max})$ for each global quantity at a representative time. In \cite{nakamura2006}, relative errors were investigated at one moment in time which, by visual inspection of Figures 2 and 3 of that work, was prior to the onset of any KH growth. \cite{pittard2009}, in contrast, present convergence results at a time well after the onset of instabilities. Convergence results in terms of the relative error $\varepsilon(N,1536)$ for the simulations are shown in Figure \ref{fig_eps_at_4}. The times depicted are $t=0.2 t_{cc}$, $t=0.4 t_{cc}$, $t=0.8 t_{cc}$, $t=1.2 t_{cc}$, $t=1.6 t_{cc}$. These plots are representative; by inspection of the quantities over the course of the simulation we find it is only very early in the simulation, $t<0.5t_{cc}$ that convergence seems apparent. After this point, there is no clear convergence in the measured quantities. Indeed, though not shown here, the behavior of the relative error varies widely over even very small timescales. The times from $0.4 t_{cc}$ to $1.6 t_{cc}$ correspond to panels $b$--$e$ in Figures~\ref{fig_1536}~and~\ref{fig_192}. The last panel at $1.6 t_{cc}$ is close in time to the comparison of Schlieren images in Fig.~\ref{fig_schlieren}.

In the first panel at $t=0.2 t_{cc}$, all the quantities except the radial velocity dispersion $\delta v_r$ show convergence in their steadily decreasing relative error with increasing resolution. The radial velocity dispersion shows an increase in relative error from \rr{96} to \rr{192}, after which it decreases. In the second panel, at $t=0.40 t_{cc}$, only clump mass $m_c$ shows monotonic decrease. The other quantities all show decreases in relative error until some resolution, at which point the relative error increased. The bottom three panels neither show monotonic decrease in relative error nor relative errors as small as those seen in the first two panels. By $0.8 t_{cc}$, the transmitted shock has not yet propagated entirely through the clump, though instabilities have been formed.

The lack of observable convergence in the bottom panels of Fig.~\ref{fig_eps_at_4} implies that either the global quantities are not a good measure of convergence after some disruption time, $t>t_{disrupt}$, or that it is inappropriate to consider convergence after $t_{disrupt}$ due to nonlinearity. It is possible that convergence is observed only before any macroscopic growth of KH or RT instabilities, or before the onset of turbulence or large-scale vorticity. Since the KH and RT growth is resolution-specific, this suggests that the small-scale instabilities do have a global effect, namely in determining the time $t_{disrupt}$. The growth rates also will be sensitive to the bow shock standoff distance: underresolving the region behind the bow shock may artificially damp the RT or KH instabilities. Thus, resolving this region is very important. The fact that cooling decreases the bow shock stand-off distance compared to adiabatic simulations implies that a higher resolution is necessary to achieve a well-converged simulation. 

\subsection{Resolving the cooling length}\label{an_coolinglength}

\begin{figure}[htbp]
\centering
\plotone{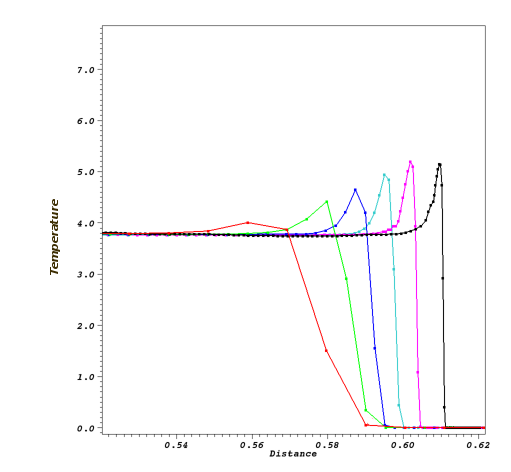}
\caption{{\small Six plots depict the on-axis temperature profile (arb. units) at time $t=0.42 t_{cc}$. Only the region near the transmitted shock is shown, and the curves have been displaced from one another for clarity. From left to right, the curves are for \rr{48}, \rr{96}, \rr{192}, \rr{384}, \rr{768}, and \rr{1536}. As resolution increases, the temperature directly behind the shock increases, and the region of cooling decreases. See \S~\ref{an_coolinglength}.}}
\label{fig_ts_temps}
\end{figure}

To get an idea of how well resolved the simulation is in the clump, we may consider the transmitted shock cooling length. We discuss the cooling length in \S~\ref{problem}, where to derive an expression for $L_{cool}$, Eq.~\ref{eq_lcool}, we assume $\Lambda\propto T^{-1/2}$. Using the expression for the cooling length in the clump, Eq.~\ref{eq_lcoolclump}, we give the approximate number of cells per cooling length in column 3 of Table~\ref{tab_parameters}.

In reality $\Lambda$ is not a simple function of temperature. From a practical standpoint, one may estimate the resolution of cooling length by visually examining the number of cells of the cooling layer behind shocks in the simulation. Figure~\ref{fig_ts_temps} shows line plots of the temperature on-axis for the simulations. Each curve has been displaced slightly to differentiate it. Simulations at \rr{12} and \rr{24} did not sufficiently resolve the cooling length and are not shown. By inspection, the number of cells which captures the decrease from $T_{max}$ behind the shock to within 10\% of $T_{ps}$ at each resolution is, \rr{48}: $1$-$2$, \rr{96}: $3$, \rr{192}: $5$, \rr{384}: $7$, \rr{768}: $11$, \rr{1536}: $14$. The apparent cooling length decreases with increasing resolution. These should be compared to the calculated values from Eq.\ref{eq_lcoolclump}, which range from 1--58 for \rr{48}--\rr{1536}. The primary cause of the discrepancy between the observed and predicted values is the deviation in the cooling curve from $\Lambda\propto T^{-1/2}$, which was assumed in Eq.~\ref{eq_lcoolclump}. 

The observed cooling length was lower than the predicted cooling length; this implies that a choice of resolution based solely on approximations like Eq.~\ref{eq_lcoolclump} may not be sufficient in a dynamic system. Without refinement criteria based on cooling, a simulation may not always adequately resolve the cooling layers behind shocks. Inadequate resolution of cooling layers increases the cooling time, effectively increasing the cooling length. Therefore, energy remains which should have been removed. On the other hand, successive shocks may drive the material into a more (or less) radiative regime, resulting in faster (slower) cooling. This confluence of effects implies that it may be difficult to recognize improperly resolved regions after the fact.

While one may choose a spatial resolution which resolves the cooling length initially, in a dynamic system this resolution may become insufficient. We propose a simple refinement criterion based on the cooling length behind shocks, $L_{cool}$, as

\begin{equation}
  N \Delta x > L_{cool} = \beta \frac{v^4}{n} 
\label{eq_criterion}
\end{equation}
where $v$ and $n$ are the velocity and number density locally, $\beta=6.61\times 10^{-11} cm^{-6}\ s^4$ from Eq.~\ref{eq_tcoolclump}, and $\Delta x$ the cell size. This therefore specifies a resolution of at least N cells per local cooling length.  Based on our experience we suggest that N must be of order 10. Other refinement criteria related to coolling have been offered, for example, \cite{niklaus2009} investigate refinement by overdensity, cooling time, and a term which depends on the enstrophy and rate of compression. Further development and comparison of refinement criteria will be illuminating.
\section{Discussion and Conclusion}\label{discussion}

``Convergence'' has a clear meaning when a problem is analytically tractable: a set of simulations are converging if an increase in resolution brings it closer to the analytic solution. Conversely, there are some cases where convergence is less well-defined, such as simulations of fully developed turbulence. For problems where there is no analytic solution, one still expects increasing resolution to improve the solution, because this for example brings the Reynolds number of the simulation closer to the actual value. (One may sidestep this by specifying a viscous scale, as in \cite{pittard2009}.) Simulations of problems with no analytic solution typically rely on a demonstration of {\it self-convergence}, which treats some high resolution simulation, believed by physical arguments to be more than sufficiently resolved, as  ``the solution.'' Lower resolution simulations in turn are compared to this ``solution,'' and a resolution is chosen which balances accuracy with computational expense.

Self-convergence for adiabatic shocked clumps shows a resolution of \rr{100}--\rr{200} is sufficient to resolve the hydrodynamics of the problem (e.g. KMC94, \cite{nakamura2006}). It is easy to see that the inclusion of other physical processes may create different resolution requirements. For example, AMR simulations involving self-gravity may choose only to base refinement criteria off of the local Jeans length \citep[e.g.][]{truelove1998}. As seen in the present work, optically thin radiative cooling prescribes scales in the problem as well which relate to the local cooling length $L_{cool}$. 

In particular, if $L_{cool}$ is highly underresolved, then wherever this is true, the simulation will be artificially moved into the isothermal regime. In \cite{mellema2002} and \cite{fragile2004}, this seems to be the case: $v_{ps}=v_s$ in the clump, where $L_{cool}/\Delta x\ll 1$ because $L_{cool}/r_{clump}\lesssim 0.1$ and there are 100--200 cells per clump radius. Thus, there is no separation between the transmitted shock and the contact discontinuity. The vanishing of this separation results in artificially damping the growth of KH and RT instabilities, analogous to the results seen in the investigation of the NTSI in \cite{kleinwoods1998} and in contrast to the results presented here, which show an increase of KH and RT growth with resolution. It will be fruitful to observe the results of repeating the above simulations, but with the cooling length correctly resolved (requiring several thousand cells per clump radius in the case of \cite{fragile2004}). The main difference between those results and ours is that they sufficiently resolved the hydrodynamics, and insufficiently resolved the cooling process, whereas in our case we progress from not resolving either process correctly to a regime in which we believe we resolve both.

Cooling enhances the growth of KH and RT instabilities (which follows from the fact that both depend on the thickness of the velocity/density shear layer, which decreases as $L_{cool}$ decreases) and, as we've seen here, this has a global effect on the simulation: the breakup and mixing of the clump proceeds differently, and measured global quantities do not agree across resolutions. There is a shorter window (proportional to $t_{KH}$ and $t_{RT}$) in which the measurement of convergence via global measured quantities will succeed. Since, without explicit viscosity, the power spectrum of vorticity is a function of resolution, once vorticies have developed one would not expect a decent correspondence between different resolutions. One must therefore stipulate an alternate criterion to ensure correct resolution of physical processes throughout the simulation, such as the criterion based on the cooling length given in Eq.~\ref{eq_criterion}. Doing so will ensure the simulation remains correctly resolved and converging toward the correct solution, even when standard measures of such convergence do not apply.

In the present suite of simulations, we seem to be approaching a converged solution. One might expect the solution to converge at \rr{3072} or \rr{6144}. However, there exists the possibility that, were the resolution to increase indefinitely, the solution would again diverge after crossing another physically-motivated resolution boundary. One could imagine several such ``plateaus'' of convergence, between which the solution appeared to diverge. It becomes all the more important, therefore, both to ensure the inclusion of correct physical processes, and to fully understand what resolution requirements are imposed by them.

\acknowledgements
The authors wish to thank Jonathan Carroll and Brandon Shroyer for useful discussions. Support for this work was in part provided by by NASA through awards issued by JPL/Caltech through Spitzer program 20269 and 051080-001, the National Science Foundation through grants AST-0507519 as well as the Space Telescope Science Institute through grants HST-AR-10972, HST-AR-11250, HST-AR-11252. KY is a recipient of the Graduate Horton Fellowship provided by the University of Rochester Laboratory for Laser Energetics. We also acknowledge funds received through the DOE Cooperative Agreement No. DE-FC03-02NA00057.

\bibliography{yirak}
\end{document}